\documentstyle[aps,prl,twocolumn,epsfig]{revtex}
\begin{document}
\draft
\title{Tau Neutrino Decays and Big Bang Nucleosynthesis}
\author{Steen Hannestad}
\address{Institute of Physics and Astronomy and Theoretical Astrophysics
Center,
University of Aarhus,
DK-8000 \AA rhus C, Denmark 
\\
and \\
Max-Planck-Institut f\"ur Physik, F\"ohringer Ring 6, 
80805 M\"unchen, Germany}
\date{\today}
\maketitle
\begin{abstract}
We investigate the non-radiative
decay during nucleosynthesis
of a massive tau neutrino with mass 0.1 - 1 MeV into an electron
neutrino and a scalar or pseudoscalar particle, $\phi$. 
The full Boltzmann equation is used and shown to give markedly
different results than the usual non-relativistic formalism for
relativistic or semi-relativistic neutrino decays. 
Indeed, the region we
investigate is where the formalism that has
previously been applied to solving this problem is expected
to break down.
We also compare the nucleosynthesis predictions from this
scenario with results from the standard model and with some
of the available observational determinations of the primordial
abundances.
It is found that for relativistic or semi-relativistic
decays the helium abundance
can be significantly lowered without changing other light
element abundances. Since a problem with the standard model
of Big Bang nucleosynthesis is that helium appears to be
overproduced, a decay of the type we discuss can be a possible
solution.

\end{abstract}


\section{Introduction}
\label{sec:Intro}
A number of indications seem to point to neutrinos having a mass. 
First of all there is the solar neutrino problem which 
is by far the strongest
evidence we have for neutrino mass \cite{solar}. 
Furthermore, there are also indications of a non-zero neutrino
mass from atmospheric neutrino data \cite{atmosphere} and,
lastly, one group claims to have seen evidence for neutrino
oscillations in a laboratory neutrino beam also indicating a 
non-zero neutrino mass \cite{lsnd}.
Now, from laboratory experiments an upper limit to the tau
neutrino mass can be obtained, which is presently $m \lesssim 24$ MeV
\cite{buskulic}.
From cosmology, we have the
well known limit on stable low mass
($m \lesssim \text{GeV}$) neutrinos \cite{kolb}
\begin{equation}
\Omega_{\nu} h^{2} = \frac{g_{\nu}}{2} 
\frac{m_{\nu}}{93.03 \text{eV}},
\label{eq:omega}
\end{equation}
using a present photon temperature of 2.736 K.
$h$ is the dimensionless Hubble constant and
$\Omega$ is the cosmological density parameter. $g_\nu =2$ for one 
flavor of neutrino and antineutrino.
Since observations demand that $\Omega_{\nu} h^{2} \leq 1$ 
\cite{kolb}, we have a mass limit on any given stable neutrino
\footnote{Note that this relation changes slightly if the heating
of neutrinos from $e^+e^-$ annihilation is included \cite{hannestad1}.}.
Thus, any neutrino with mass in the range 100 eV - 24 MeV is
necessarily unstable.
There are, however, 
many possible modes of decay for a
massive neutrino. 

For example there is the predicted
decay \cite{kim}
$\nu_{\i} \to \nu_{j}e^+e^-$
if the mass is larger than $2 m_{e}$ and the mixing angle between the
two neutrinos is different from zero. A flavour changing neutral
current can also lead to the decay $\nu_{i} \to \nu_{j}
\overline{\nu_{j}}\nu_j$.
There can also be other
more exotic modes of decay, for example decay via emission
of scalars or pseudo-scalars. This decay mode is generic for example
in the majoron models of neutrino mass \cite{majo1}.

The effect of such unstable tau neutrinos on nucleosynthesis have
been investigated many times in the literature
\cite{terasawa,kands,sandturn,madsen,gyuk,kawasaki,kawasaki2},
the most recent investigations
being those of Dodelson, Gyuk and Turner \cite{gyuk} 
and Kawasaki et {\it al.} 
\cite{kawasaki,kawasaki2}.
Dodelson, Gyuk and Turner have performed a 
detailed study of several possible decay modes in the context of
non-relativistic decays, whereas Kawasaki et {\it al.} have performed a 
calculation using the full Boltzmann equation for the decay mode
$\nu_{\tau} \to \nu_{\mu} \phi$ \cite{majo2,majo3}.
In all cases it is found that it is possible to change significantly the
primordial abundances via decay of the tau neutrino.

In the present paper we focus on the decay 
\begin{equation}
\nu_{\tau} \to \nu_e \phi,
\end{equation}
where $\nu_{\tau}$ is assumed to be a Majorana particle and $\phi$ is
a light scalar or pseudoscalar particle.
This differs from the decay $\nu_{\tau} \to \nu_{\mu} \phi$ in that
it includes an electron neutrino in the final state. Since $\nu_e$
enters directly into the weak interactions that interconvert
neutrons and protons this decay can potentially alter the
outcome of nucleosynthesis drastically.
Indeed the non-relativistic results of Dodelson, Gyuk and Turner indicate
that the primordial helium abundance, 
Y$_P$, can be changed radically, either
increasing or decreasing Y$_P$ depending on the mass and lifetime of
the tau neutrino.

Now, in the last few years, evidence has been gathering that 
the standard picture of the way light nuclei are formed
in the early Universe may be facing a crisis
\cite{hata}.
The main point is that helium is overproduced relative to the
other light nuclei so that the standard theoretical predictions
are only marginally consistent with the observational results
\cite{hata}.
Other measurements of the primordial helium abundance do yield somewhat
higher values \cite{isotov},
and the unknown systematical errors both in observations
and in chemical evolution calculations may, however, be larger than
presently assumed so that it is perhaps premature to talk of
a real ``crisis'' for Big Bang nucleosynthesis.

Our approach will not be so much to discuss the specific limits
from nucleosynthesis since these are still quite uncertain
as it will be a discussion of the differences between our way of
solving the Boltzmann equations and those previously used.
Nevertheless, in light of the possibility that some new element
is missing from the standard nucleosynthesis calculations
we think that it is important to try and find methods
of changing the nucleosynthesis predictions by including
plausible new physics in the calculations.
One possible way of doing this is to include a massive and unstable
tau neutrino.

In order to obtain good fits to the observational data it 
is, as just mentioned,
necessary to lower the helium abundance somewhat compared to
the other light nuclei.
This can be achieved by having relatively low mass tau neutrinos
decay while they are still relativistic or semi-relativistic.
However, this is exactly the region where the 
non-relativistic formalism
breaks down because it assumes a delta function momentum
distribution of the decay products and neglects
inverse decays. It is therefore of significant
interest to investigate this decay using the full Boltzmann
equation in order to calculate abundances in this parameter
region.

In the present paper we calculate the expected primordial abundances
for a tau neutrino mass in the range 0.1 - 1 MeV.
In Section II we describe the necessary formalism needed for this
calculation. In Section III we discuss our numerical results.
Section IV contains a description of our nucleosynthesis 
calculations compared to the observational data and finally 
Section V contains a summary and discussion.

\section{Necessary formalism}
\label{sec:Forma}
The fundamental way to describe the evolution of different particle
species in the early Universe is to use the Boltzmann equation
\begin{equation}
L[f] = \sum C_{i}[f],
\end{equation}
where the sum is over different possible collisional terms for the
given particle, such as decay, scattering and pair-annihilation.
In our case, we include the standard weak interactions of neutrinos
with each other and with electrons and positrons. Furthermore we
include a decay term. We shall assume, however, that the scalar
particles are collisionless except for the decays and inverse 
decays. That is, they have no self interactions and no other
interactions with neutrinos. This may or may not be a good 
assumption, depending on the various coupling constants. It greatly
simplifies the calculations, however.
Now, the various terms in the Boltzmann equation can be written 
as follows
\begin{equation}
L[f] = \frac{\partial f}{\partial t}-\frac{\text{d} R}
{\text{d} t}\frac{1}{R}p\frac{\partial f}{\partial p},
\end{equation}

Since there are only 2-particle interactions like $1+ 2 \rightarrow 
3 +4$, $C_{\text{weak}}$ can be written as

\begin{eqnarray}
C_{\text{weak}}[f] & = & \frac{1}{2E_{1}}\int d^{3}\tilde{p}_{2}
d^{3}\tilde{p}_{3}d^{3}\tilde{p}_{4}
\Lambda(f_{1},f_{2},f_{3},f_{4})\times 
\label{integral}\\ 
& & S \sum \! \mid \! M \! \mid^{2}_{12\rightarrow 34}\delta^{4}
({\it p}_{1}+{\it p}_{2}-{\it p}_{3}-{\it p}_{4})(2\pi)^{4}.
\nonumber
\end{eqnarray}
where \(\Lambda(f_{1},f_{2},f_{3},f_{4})=(1-f_{1})(1-f_{2})f_{3}
f_{4}-(1-f_{3})(1-f_{4})f_{1}f_{2}\) is the phase space factor, 
including Pauli blocking of the final states, and $d^{3}\tilde{p}
 = d^{3}p/((2 \pi)^{3} 2 E)$. $S$ is a symmetrization
factor of 1/2! for each pair of identical particles in initial or 
final states \cite{wagoner},
and $\sum \! \mid\!\!M\!\!\mid^{2}$ is the weak interaction matrix element
squared and spin-summed. The matrix elements for the relevant
processes have been compiled for example by Hannestad and Madsen
\cite{hannestad}.
${\it p}_{i}$ is the four-momentum of particle $i$.

Since we are only looking at Majorana neutrinos the decay terms are 
quite simple. Since there is almost no net lepton number in the early
Universe the Majorana neutrino is effectively an unpolarised
species. However, this means that there can be no preferred direction
in the rest frame of the parent particle. Therefore the decay is
necessarily isotropic in this reference frame. In this case the decay
terms can be written as \cite{starkman}

\begin{eqnarray}
C_{\text{dec}}[f_{\nu_{\tau}}] & = &
- \frac{m_{\nu_{\tau}}^{2}}{\tau m_{0} E_{\nu_{\tau}}
p_{\nu_{\tau}}}
\int_{E_{\phi}^{-}}^{{E_{\phi}^{+}}}dE_{\phi}
\Lambda(f_{\nu_{\tau}},f_{{\nu_{e}}},f_{\phi})
\end{eqnarray}

\begin{eqnarray}
C_{\mbox{\scriptsize{dec}}}[f_{{\nu_{e}}}] & = &
\frac{g_{\nu_{\tau}}}{g_{{\nu_{e}}}} \frac{m_{\nu_{\tau}}^{2}}
{\tau m_{0} E_{{\nu_{e}}} 
p_{{\nu_{e}}}}
\int_{E_{\nu_{\tau}}^{-}}^{{E_{\nu_{\tau}}^{+}}}dE_{\nu_{\tau}} 
\Lambda(f_{\nu_{\tau}},f_{{\nu_{e}}},f_{\phi})
\end{eqnarray}

\begin{eqnarray}
C_{\mbox{\scriptsize{dec}}}[f_{\phi}] & = &
\frac{g_{\nu_{\tau}}}{g_{\phi}}
\frac{m_{\nu_{\tau}}^{2}}{\tau m_{0} E_{\phi} 
p_{\phi}}
\int_{E_{\nu_{\tau}}^{-}}^{{E_{\nu_{\tau}}^{+}}}dE_{\nu_{\tau}} 
\Lambda(f_{\nu_{\tau}},f_{{\nu_{e}}},f_{\phi}),
\label{bosdec}
\end{eqnarray}
where
$\Lambda(f_{{\nu_{\tau}}},f_{{\nu_{e}}},f_{\phi}) = f_{{\nu_{\tau}}}
(1-f_{{\nu_{e}}})(1+f_{\phi})-
f_{{\nu_{e}}}f_{\phi}(1-f_{{\nu_{\tau}}})$, 
$m_{0}^{2} = m_{{\nu_{\tau}}}^{2}-2(m_{\phi}^{2}+m_{{\nu_{e}}}^{2})+
(m_{\phi}^{2}-m_{{\nu_{e}}}^{2})^{2}/m_{{\nu_{\tau}}}^{2}$.
$\tau$ is the lifetime of the heavy neutrino and $g$ is the
statistical weight of a given particle. We use $g_{{\nu_{\tau}}} = 
g_{{\nu_{e}}} = 2$ and
$g_{\phi} = 1$, corresponding to $\phi = \overline{\phi}$. This
assumption is not significant to the present investigation.
Furthermore we shall assume that the masses of $\nu_e$ and $\phi$
are effectively zero during nucleosynthesis.

The integration limits are
\begin{eqnarray}
E_{{\nu_{\tau}}}^{\pm} (E_{i}) & = & \frac{m_{0}m_{{\nu_{\tau}}}}
{2m_{i}^{2}}[E_{i}(1+4(m_{i}/m_{0})^{2})^{1/2} \pm \\ \nonumber
& & (E_{i}^{2}-m_{i}^{2})^{1/2}]
\end{eqnarray}
and
\begin{equation}
E_{i}^{\pm} (E_{{\nu_{\tau}}}) = \frac{m_{0}}{2m_{H}}
[E_{{\nu_{\tau}}} (1+4(m_{i}/m_{0})^{2})^{1/2} \pm p_{{\nu_{\tau}}}]
\label{energyi}
\end{equation}
where the index $i = {\nu_{e}},\phi$.

Apart from the Boltzmann equation one needs equations to relate
the evolution of 
time, the cosmic expansion rate and the photon temperature. These
quantities can
be calculated by use of the energy conservation equation
\begin{equation}
{d}(\rho R^{3})/dt + p {d}(R^{3})/dt = 0
\label{conserve}
\end{equation}
and the Friedmann equation
\begin{equation}
H^{2} = 8 \pi G \rho/3.
\label{friedmann}
\end{equation}
$R$ is the cosmological scale factor, $H$ is the Hubble parameter and
$\rho$ is the total energy density of all particles present.

\section{Numerical Results}
\label{sec:Results}

We have solved the Boltzmann equation for the evolution of distribution
functions together with the energy conservation equation
, Eq.\ (\ref{conserve}),
and the Friedmann equation, Eq.\ (\ref{friedmann}).
Specifically we have solved for masses of 0.1-1 MeV and lifetimes
larger than 0.1 s.

\begin{figure}[t]
\begin{center}
\epsfysize=7truecm\epsfbox{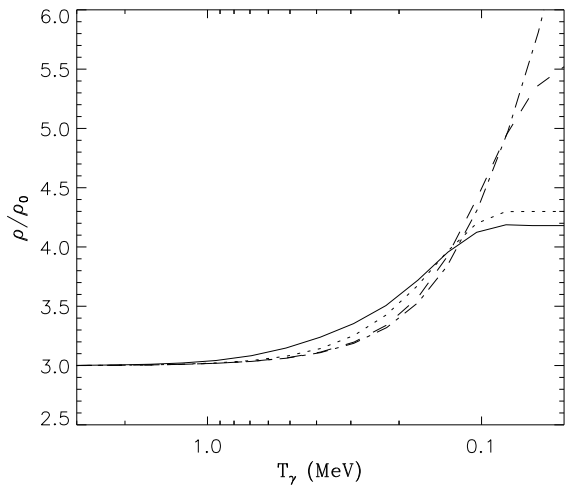}
\vspace{0truecm}
\end{center}
\baselineskip 17pt

\label{fig1}
\caption{The energy density of the different neutrinos and the
scalar particle in units of the energy density of a standard
massless neutrino for a tau neutrino mass of 0.5 MeV.
The full line is for $\tau = 1$ s, the dotted for
$\tau = 10$ s, the dashed for $\tau = 100$ s and the
dot-dashed for $\tau = 1000$ s.}
\end{figure}

In Fig. 1 we show the evolution of energy density in neutrinos and
the pseudoscalar particle for a tau neutrino mass of 0.5 MeV.
The energy density evolves quite differently in the different cases.
Since the energy density in a non-relativistic species only 
decreases as $R^{-3}$ compared to $R^{-4}$ for relativistic particles
the rest mass energy of the tau neutrino will dominate completely
at late times if it is stable. 
If it decays the rest mass energy is transferred into relativistic
energy so that the total energy density no longer increases
relative to that of a single standard massless neutrino species.
This difference is clearly seen
between different tau neutrino lifetimes.

In Fig. 2 we show the spectral distribution of the electron neutrino
for a tau neutrino mass of 0.5 MeV and different lifetimes. To understand
this plot better
we can define a ``relativity parameter'', $\mu$,
 for the decay
\begin{equation}
\mu_{\nu_{\tau}} \equiv \frac{m_{\nu_{\tau}}^{2} 
\tau_{\nu_{\tau}}}{9 \text{MeV}^{2} \text{s}}.
\end{equation}
A particle shifts from relativistic to 
non-relativistic at a temperature of roughly $T \simeq m/3$.
When the Universe is radiation dominated
\begin{equation}
\frac{t}{1 \text{s}} \simeq \left(\frac{T}{1 \text{MeV}}\right)^{-2}.
\end{equation}

\begin{figure}[t]
\begin{center}
\epsfysize=7truecm\epsfbox{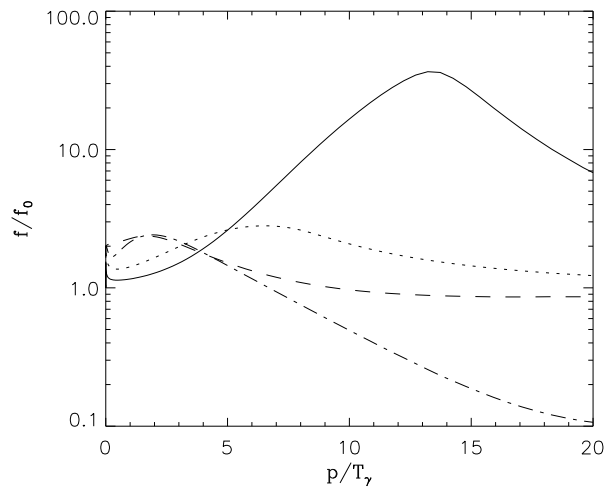}
\vspace{0truecm}
\end{center}
\baselineskip 17pt

\caption{The electron neutrino distribution at asymptotically 
low temperature (after complete decay) in units of the distribution
of a standard massless neutrino. The tau neutrino mass is 0.5 MeV.
The full line is for $\tau = 1$ s, the dotted for
$\tau = 10$ s, the dashed for $\tau = 100$ s and the
dot-dashed for $\tau = 1000$ s.}
\label{fig2}
\end{figure}
Therefore, if the decay is relativistic, 
\begin{equation}
\tau < t(T = m/3) \simeq \frac{9 m^{-2}}{\text{MeV}^{-2}} \text{s}.
\end{equation}
Thus, if $\mu_{i} < 1$ the decay is relativistic, whereas if 
$\mu_{i} > 1$ it is non-relativistic.
For lifetimes of 1, 10, 100 and 1000 s the relativity parameters
are respectively 0.028, 0.28, 2.78 and 27.8.
For non-relativistic decays the decay neutrino distribution
assumes a rather narrow shape coming from the delta function
energy distribution.
For very short lifetimes the decay installs an equilibrium
between $\nu_e$, $\nu_{\tau}$ and $\phi$ because of rapid
inverse decays. This can lead to a
significant depletion of high momentum electron neutrinos as also
noted by Madsen \cite{madsen} who treated this case of very short
lifetimes using equilibrium thermodynamics.

\section{nucleosynthesis effects}

In order to estimate the effect on nucleosynthesis, we have
employed the nucleosynthesis code of Kawano \cite{kawano}, modified in order 
to incorporate a decaying neutrino. This includes taking into account
the changing energy density as well as the change in electron neutrino
distribution.

A decaying tau neutrino can affect nucleosynthesis in several 
different ways. Firstly, the cosmic energy density $\rho$ is
changed. Since the cosmic expansion rate is given directly in terms
of this energy density via the Friedmann equation, 
Eq.\ (\ref{friedmann}), it is also
changed. It is a well known fact that increasing the energy
density leads to an earlier freeze-out of the n-p conversion
and therefore produces more helium \cite{kolb},
whereas decreasing the energy density decreases the helium fraction.
This is the effect discussed by Kawasaki et {\it al.}  
\cite{kawasaki2}, namely that
an MeV neutrino decaying into sterile daughter products while
still relativistic or semirelativistic can actually decrease 
the cosmic energy density thereby decreasing the helium abundance.  

However, there is also another another effect stemming from the
change in electron neutrino temperature. Since the electron neutrino
enters directly into the n-p processes this can be called a ``first
order'' effect and is potentially much more important than the
``second order'' effect of changing the energy density.
This effect of change in the electron
neutrino temperature has already been discussed 
by several authors in the context
of non-relativistic decays \cite{terasawa,gyuk}.

\begin{figure}[t]
\begin{center}
\epsfysize=7truecm\epsfbox{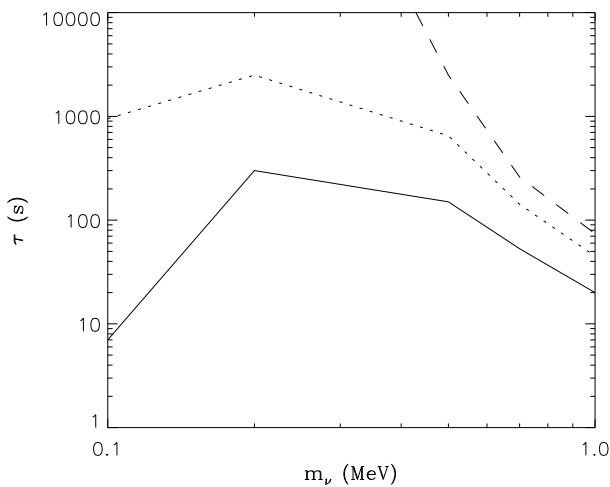}
\vspace{0truecm}
\end{center}
\baselineskip 17pt

\caption{Helium abundance contours as a function of tau
neutrino mass and lifetime for a baryon-to-photon ratio, $\eta$,
of $3 \times 10^{-10}$. The full line is Y$_{P}$ = 0.20,
the dotted is Y$_{P}$ = 0.22 and the dashed is Y$_{P}$ = 0.24.
The value in the standard model for this $\eta$ is Y$_{P}$ = 0.2389.}
\label{fig3}
\end{figure}
If the decay is non-relativistic, the energy of a produced electron
neutrino is $m$/2. If this energy is significantly above the
energy threshold for the two processes
\begin{equation}
p + \overline{\nu_e} \to n + e^+
\end{equation}
and
\begin{equation}
p + \overline{\nu_e} + e^- \to n,
\end{equation}
the decaying tau neutrinos
will act to produce more He \cite{gyuk,terasawa}. The reason is
that the absorption cross section at high energies is the same
on neutrons and protons.
Since there are many more protons present than neutrons,
more neutrons will be produced. In the end this leads to
a higher helium fraction.
However, if the mass of the decaying neutrino is below this
threshold the produced electron neutrinos will stimulate the
conversion of neutrons into protons thereby actually decreasing
the He abundance. This effect then competes with the rest mass
effect which increases Y$_P$.

If the decay is relativistic the electron 
neutrinos are produced at roughly thermal 
energies. Effectively this amounts to increasing the electron
neutrino temperature. This in turn leads to a decrease in helium
production. If the decay takes place at high temperatures it is
because beta equilibrium is kept for a longer time, whereas if the
decay takes place at temperatures below the threshold for proton
to neutron conversion it still leads to lower helium abundance because
an increase in the electron neutrino temperature stimulates
the conversion of neutrons to protons over the inverse
reaction.

In Fig. 3 we show contour lines for the helium abundance as
function of neutrino mass and lifetime. 
It is seen that
helium can be significantly suppressed relative to the standard
case if the lifetime is short enough and increased if the mass and
lifetime are both high. If the mass and lifetime are both
sufficiently high the helium abundance is instead increased.
Notice also that even for small masses of the order 0.1 MeV the
helium abundance can be changed significantly compared to the
standard value for rather a large range of lifetimes. 

Our Fig. 3 should be compared with for example the results
of Terasawa and Sato \cite{terasawa} or Dodelson, Gyuk and Turner \cite{gyuk}
obtained using non-relativistic theory.
The most straightforward comparison is with Figs. 2b and
3b in Ref. \cite{terasawa}. For long lifetimes the difference is
quite small as would be expected since this is the non-relativistic
limit. However, for short lifetimes the difference is significant.
For very short lifetimes, our calculated He abundance is larger
than that of Terasawa and Sato. The reason is that if one uses
the full Boltzmann equation in this case, decays and inverse
decays will bring the particle distributions into equilibrium
as discussed in Sec.~3.
Thus, if one keeps on going to shorter and shorter lifetimes nothing
new happens since it is already decay in equilibrium. Therefore
our curve for He flattens out instead of decreasing for very short
lifetimes.
For somewhat longer lifetimes our He abundance is on the other
hand smaller than that found by Terasawa and Sato. The reason here is
that the decay produces a peak of very low momentum electron
neutrinos and that these states are not upscattered because
the weak interactions have already frozen out. In the end this produces
a somewhat colder electron neutrino distribution than would have
been obtained using non-relativistic theory and therefore predicts
less helium.
This low momentum peak can be seen in Fig.~2 for the example of
a 0.5 MeV $\tau$ neutrino. For non-relativistic decays this peak
disappears because low momemtum states are not energetically
accessible.
In essence our predicted curve for the He abundance is therefore much
flatter at short or intermediate lifetimes than what one would obtain
using the non-relativistic formalism. For very long lifetimes our
calculation fits fairly well with that obtained by Terasawa and Sato
as could be expected.

\begin{figure}[t]
\begin{center}
\epsfysize=7truecm\epsfbox{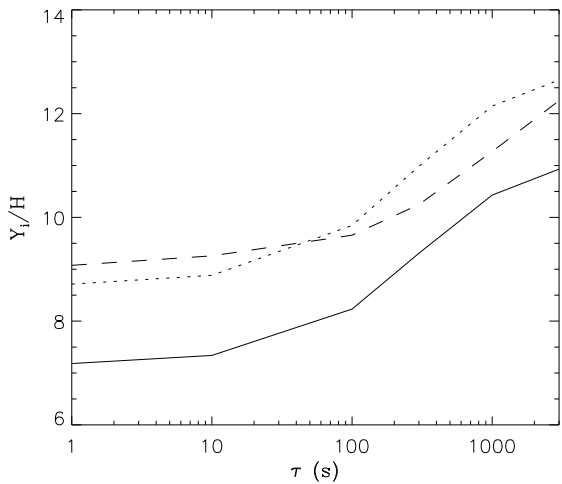}
\vspace{0truecm}
\end{center}
\baselineskip 17pt

\caption{The abundance of D, $^3$He and $^7$Li as a function
of tau neutrino lifetime. The curves have been calculated for
$m = 0.5$ MeV and  $\eta = 3 \times 10^{-10}$. The full line shows
$(\text{D/H})/10^{-5}$ the dashed shows $(\text{(D+$^3$He)/H})/10^{-5}$
and the dot-dashed shows $(\text{$^7$Li/H})/10^{-11}$.}
\label{fig4}
\end{figure}

In Fig. 4 we show the abundance of D, $^3$He and $^7$Li for a
specific example of m = 0.5 MeV. We see that the abundances of these
elements only change by relatively small amounts even for great variations
in neutrino lifetime.
Thus, the main effect of the decay is to lower the helium abundance 
while leaving the other abundances more or less unchanged.

The calculated abundances for different masses and lifetimes are
compared with observational limits. Unfortunately there is a great
deal of controversy connected with these. However, since our main
emphasis is on the differences between our approach 
to solving the Boltzmann equations and the non-relativistic
approximations previously used, and not 
so much on the specific
nucleosynthesis limits to tau neutrino mass and lifetime we will
not go into too much detail regarding this point.
For $^4$He we use the value calculated by
Hata et {\it al.} \cite{hata} of
\begin{equation}
{\rm Y}_P = 0.232 \pm 0.002 \pm 0.005.
\end{equation}
For deuterium the situation is somewhat complicated. From measurements
in the local interstellar medium one can obtain a deuterium
abundance of \cite{hata}
\begin{equation}
\text{D/H} \simeq  1.6 \times 10^{-5},
\end{equation}
which can be viewed as a lower limit to the primordial abundance.
However, some recent results from QSO absorption systems seem
to indicate a primordial value much higher than this \cite{songaila}
\begin{equation}
\text{D/H} \simeq  1.9 - 2.5 \times 10^{-4}.
\end{equation}
Other similar observations yield much lower values, closer to the
local one \cite{tytler}.
In light of the controversy of using deuterium results from 
these measurements, we use the locally obtainable lower limit
in the present paper.
From evolution arguments one can also obtain an upper limit to the
primordial D+$^3$He abundance of \cite{copi}
\begin{equation}
\text{(D+$^3$He)/H} \lesssim  1.1 \times 10^{-4}.
\end{equation}
Finally for the abundance of $^7$Li we use a bound of
\begin{equation}
\text{$^7$Li/H} =  1.4 \pm 0.3 ^{+1.8}_{-0.4} \times 10^{-10}
\end{equation}
obtained by Copi, Schramm and Turner \cite{copi}.
\begin{figure}[t]
\begin{center}
\epsfysize=7truecm\epsfbox{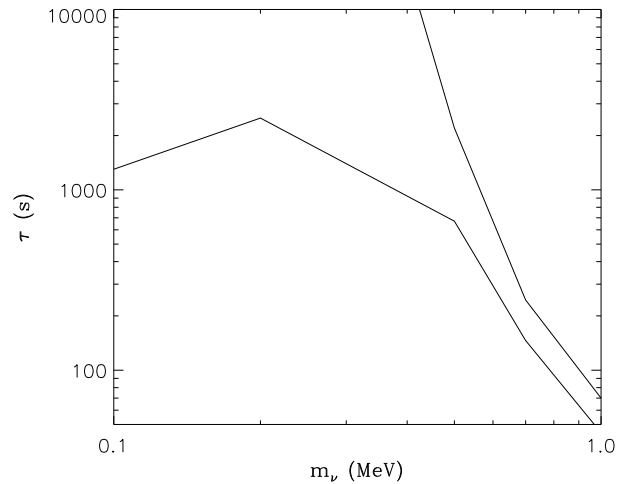}
\vspace{0truecm}
\end{center}
\baselineskip 17pt

\caption{Allowed region of tau neutrino mass and lifetime. The
allowed region is between the two full lines.}
\label{fig5}
\end{figure}

Altogether these are the observational values which the theoretical
predictions should be able to reproduce. 
In the standard model the theoretical 
predictions are only marginally consistent with observations
because helium is overproduced compared to the other light
nuclei.
In our scenario this problem is resolved by having the tau neutrino
decay during nucleosynthesis into an electron neutrino final state.

In Fig. 5 we show the allowed region of lifetime versus mass for
the tau neutrino using the above constraints.
In all the mass interval from 0.1-1 MeV it is possible to obtain
a fit to the observed abundances. Note however, that for masses
in the high end of this region a fit can only be obtained in a very
narrow lifetime interval. This is because of the very steep 
dependence of $Y_P$ on the lifetime in this region.
For lower masses a good fit can be obtained in a broad region of 
lifetimes.

Another important fact is that since the helium abundance
is lowered without disturbing greatly the other abundances, the
upper and lower bound on the baryon-to-photon ratio, $\eta$, is now given
essentially only by the limits coming from D, $^3$He and $^7$Li.
This also means that a relatively high value for $\eta$ can be 
accommodated, about $6 \times 10^{-10}$, coming from requiring that
$^7$Li should not be overproduced.

In Fig.~6 we show the allowed region of $\eta_{10}$ as a function
of tau neutrino lifetime for three different masses. We have also
plotted the upper and lower limits to $\eta_{10}$ from the
standard calculation.
\begin{figure}[t]
\begin{center}
\epsfysize=7truecm\epsfbox{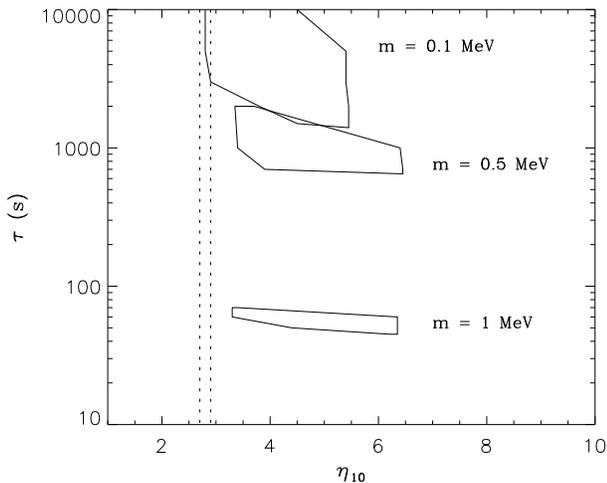}
\vspace{0truecm}
\end{center}
\baselineskip 17pt

\caption{Allowed regions of $\eta_{10} \equiv 10^{10} \times \eta$ 
for different tau neutrino
masses and lifetimes. The regions inside the full lines are allowed
regions. The vertical dotted lines show the consistency interval
for  $\eta_{10}$ in the standard model for our chosen observational
constraints.}
\label{fig6}
\end{figure}

\section{conclusion}

We have studied the decay of a relatively low mass
 tau neutrino into an
electron neutrino and a scalar or pseudoscalar particle using
the full Boltzmann equation.
It was found that the primordial helium abundance, Y$_P$, can 
change drastically compared to the standard value.
This is in concordance with the findings of previous
authors who used a non-relativistic treatment
\cite{terasawa,gyuk}. Our actual numerical values differ 
significantly from those previously obtained by use
of non-relativistic formalism, but the
general trend is the same, namely that low mass neutrinos
decaying while relativistic or semi-relativistic lower the helium
abundance.

The decay we have studied differs completely from the 
$\nu_{\tau} \to \nu_{\mu} \phi$ decay studied by Kawasaki et {\it al.}
\cite{kawasaki,kawasaki2}
because the electron neutrinos
directly affect the weak reaction rates that interconvert
neutrons and protons.
Only if much more reliable estimates of the primordial abundances
are developed will it be possible to discern between the 
two different decay modes.

Our aim has mainly been to discuss the differences between using the
full Boltzmann formalism and using the non-relativistic 
approximation in doing these calculations. We have
not done very detailed statistical analysis in order to obtain
strict nucleosynthesis limits.

However, it was shown that a good fit to the observed primordial abundances
can be achieved for a large range of different masses and lifetimes.
Given the possibly large unknown systematical errors
in the observations and chemical evolution models it is 
perhaps too early to talk of a real crisis for
Big Bang nucleosynthesis. However, once the observational
bounds become more strict there might very well turn out to be
such a crisis.

In light of this possible discrepancy 
between observed and predicted
abundances we still feel that 
it is important to explore possible ways to
change the light element abundances via plausible introduction of
new physics. The tau neutrino decay into an electron neutrino
final state is just such a possibility.

Perhaps one should also finally note that even if the helium
abundance turns out to have been significantly underestimated
a tau neutrino decay of the type we have discussed can still
make nucleosynthesis predictions fit the observations, but for
completely different values of mass and lifetime.

\acknowledgements{This paper has been supported by a grant from the
Theoretical Astrophysics Center under the Danish National Research
Foundation. I wish to thank the Max Planck Institute for
Physics in Munich for their hospitality during the writing of this
paper as well as Georg Raffelt and Jes Madsen who have provided 
constructive discussion and criticism.}


\begin{references}
\bibitem{solar}see for example J. Bahcall, 
talk given at 18th Texas Symposium on
Relativistic Astrophysics, Chicago IL, Report no. astro-ph/9702057.
\bibitem{atmosphere}See for example T. K. Gaisser, Talk given at the 17th 
International Conference on Neutrino Physics and Astrophysics
(Neutrino 96), Helsinki, Finland, Report no. hep-ph/9611301.
\bibitem{lsnd}C. Athanassopoulos et {\it al.},
Phys.\ Rev.\ Lett.\ {\bf 77}, 3082 (1996). 
\bibitem{buskulic}D. Buskulic et {\it al.}, Phys.\ Lett.\ B {\bf 349},
585 (1995).
\bibitem{kolb} E.W. Kolb and M.S. Turner, {\it The Early Universe}, 
Addison Wesley (1990).
\bibitem{hannestad1}S. Hannestad and J. Madsen,
Phys.\ Rev.\ D\ {\bf 52}, 1764 (1995). 
\bibitem{kim}See for example 
C. W. Kim and A. Pevsner, {\it Neutrinos in Physics 
and Astrophysics}, Harwood Academic Publishers (1993).
\bibitem{majo1}Y. Chikashige, R. Mohapatra and R. Peccei,
Phys.\ Rev.\ Lett.\ {\bf 45}, 1926 (1980); G. Gelmini and M. Roncadelli,
Phys.\ Lett.\ {\bf B99}, 411 (1981). For a good textbook review
of the different majoron models see R.\ N.\ Mohapatra and P.\ B.\
Pal, {\it Massive Neutrinos in Physics and Astrophysics},
World Scientific Lecture Notes in Physics - Vol. 41, World Scientific
1991. 
\bibitem{terasawa}N. Terasawa and K. Sato, Phys.\ Lett.\ B\
{\bf 185}, 412 (1987).
\bibitem{kands}E. W. Kolb and R. J. Scherrer,  
Phys.\ Rev.\ D\ {\bf 25}, 1481 (1982). 
\bibitem{sandturn}R. J. Scherrer and M. S. Turner, Astrophys.\ J.\
{\bf 331}, 19 (1988); R. J. Scherrer and M. S. Turner {\it ibid.} 
{\bf 331}, 31 (1988).
\bibitem{madsen}J. Madsen, Phys.\ Rev.\ Lett.\
{\bf 69}, 571 (1992).
\bibitem{gyuk}S. Dodelson, G. Gyuk and M.S. Turner, 
Phys.\ Rev.\ D\ {\bf 49}, 5068 (1994). 
\bibitem{kawasaki}M. Kawasaki et {\it al.}, Nucl.\ Phys.\ {\bf B419},
105 (1994).
\bibitem{kawasaki2}M. Kawasaki, K. Kohri and K. Sato,
Report no. astro-ph/9705148.
\bibitem{majo2}This type of decay has been discussed several times
in the literature. For earlier papers on the subject, see for example
J.\ W.\ F.\ Valle, Phys.\ Lett.\ {\bf B131}, 87 (1983);
J.\ Schechter and J.\ W.\ F.\ Valle, Phys.\ Rev.\ D {\bf 25}, 
774 (1982); G.\ Gelmini and J.\ W.\ F.\ Valle,
Phys.\ Lett.\ {\bf B142}, 181 (1984).
\bibitem{majo3}Another related issue which has discussed 
is the possibility of changing nucleosynthesis
through the pair annihilation of massive $\tau$ neutrinos
into majorons, see A.\ D.\ Dolgov et {\it al.}, 
Nucl.\ Phys.\ {\bf B496}, 24 (1997) and references therein.
\bibitem{hata}N. Hata et {\it al.}, Phys.\ Rev.\ Lett.\
{\bf 75}, 3977 (1995).
\bibitem{isotov}Y. I. Izotov, T. X. Thuan and V. A. Lipovetsky,
Astrophys. J. Suppl. {\bf 108}, 1 (1997). 
\bibitem{wagoner}R. V. Wagoner, in Physical Cosmology - Les Houches 
(1979), eds. R. Balian, J. Audouze and D. N. Schramm.
\bibitem{hannestad}S. Hannestad and J. Madsen, Phys.\ Rev.\ D\
{\bf 54} 7894 (1996).
\bibitem{starkman}G. D. Starkman, N. Kaiser and R. A. Malaney, 
Astrophys. J. {\bf 434}, 12 (1994).
\bibitem{kawano}L. Kawano, Report no. FERMILAB-Pub-92/04-A (1992)
(unpublished).
\bibitem{songaila}A. Songaila et {\it al.}, Nature (London)
{\bf 368}, 599 (1994).
\bibitem{tytler}D. Tytler, S. Burles and D. Kirkman, Report no.
astro-ph/9612121.
\bibitem{copi}C. J. Copi, D. N. Schramm, and M. S. Turner, 
FERMILAB-Pub-94/174-A (1994).



\end{references}
\end{document}